# Electromagnetic Structure of Mesons in a Light-Front Constituent Quark Model

F. Cardarelli [1], I.L. Grach [2], I.M. Narodetskii [2], G. Salmè [3], S. Simula [3]

[1] Istituto Nazionale di Fisica Nucleare, Sezione Tor Vergata, Via della Ricerca Scientifica 1, I-00133 Roma, Italy
[2] Institute for Theoretical and Experimental Physics, 117259 Moscow, Russia
[3] Istituto Nazionale di Fisica Nucleare, Sezione Sanità, Viale Regina Elena 299, I-00161 Roma, Italy

**Abstract.** The elastic and transition electromagnetic form factors of mesons are investigated within a light-front constituent quark model. The general formulae for the space-like matrix elements of the one-body component of the electromagnetic current, including both Dirac and Pauli form factors of the constituent quarks, are presented. Our results for the pion charge form factor and the $\pi\rho$ transition form factor are reported and compared with those obtained within various relativistic approaches in a range of values of the four-momentum transfer accessible to $CEBAF$.

The investigation of elastic and transition electromagnetic (e.m.) form factors of mesons has recently received a renewed interest, because measurements of the pion and kaon charge form factors, as well as of the electroproduction cross section of vector mesons off the nucleon, are planned at $CEBAF$ [1]. The aim of this contribution is to address the calculation of meson e.m. form factors, adopting the constituent quark ($CQ$) model for the description of the meson structure and the Hamiltonian light-front formalism [2] for a Poincaré-covariant treatment of the $CQ$ degrees of freedom. For space-like values of the four-momentum transfer, all the relevant formulae needed for the calculation of the matrix elements of the one-body component of the e.m. current, including both Dirac and Pauli form factors for the $CQ$'s, will be presented. Moreover, the results of our calculations of the pion charge form factor and of the $\pi\rho$ transition form factor will be reported and compared with those obtained from various relativistic approaches at momentum transfers accessible to $CEBAF$.

As is well known (cf. [2]), the light-front formalism allows an exact separation in momentum space between the center of mass and intrisic wave functions; therefore, in what follows, we will focus on the intrisic part of the meson wave function. Let us remind that the intrinsic light-front kinematical variables are $\boldsymbol{k}_\perp = \boldsymbol{p}_{q\perp} - \xi \boldsymbol{P}_\perp$ and $\xi = p_q^+/P^+$, where the subscript $\perp$ indicates the projection perpendicular to the spin quantization axis, defined by the vector $\hat{n} = (0, 0, 1)$, and the *plus* component of a 4-vector $p \equiv (p^0, \boldsymbol{p})$ is given by $p^+ = p^0 + \hat{n} \cdot \boldsymbol{p}$; eventually, $\tilde{P} \equiv (P^+, \boldsymbol{P}_\perp) = \tilde{p}_q + \tilde{p}_{\bar{q}}$ is the total light-front momentum of the meson. Omitting for simplicity the colour degrees of freedom, the requirement of Poincaré covariance for the intrinsic wave function $\langle \nu \bar{\nu} | JM \rangle$ of a meson with angular momentum $J$ and projection $M$ implies

$$\langle \nu \bar{\nu} | JM \rangle = \sqrt{A(\boldsymbol{k}_\perp, \xi)} \sum_{SM_S} R^{SM_S}(\xi, \boldsymbol{k}_\perp; \nu \bar{\nu}) \, \tilde{w}^{JM}_{SM_S}(\boldsymbol{k}) \qquad (1)$$

where $\nu, \bar{\nu}$ are the $CQ$ spin variables; $k^2 \equiv k_\perp^2 + k_n^2$; $k_n \equiv (\xi - 1/2)M_0 + (m_{\bar{q}}^2 - m_q^2)/2M_0$; $M_0^2 = (m_q^2 + k_\perp^2)/\xi + (m_{\bar{q}}^2 + k_\perp^2)/(1-\xi)$; $A(\boldsymbol{k}_\perp, \xi) = M_0[1 - (m_q^2 - m_{\bar{q}}^2)^2/M_0^4]/4\xi(1-\xi)$. In Eq. (1) $\tilde{w}^{JM}_{SM_S}(\boldsymbol{k}) = \sum_{LM_L} w^{q\bar{q}}_{LSJ}(k^2) \, Y_{LM_L}(\hat{k}) \, \langle LM_L SM_S | JM \rangle$, where $L$ is the orbital angular momentum, $Y_{LM_L}$ the usual spherical harmonics and $w^{q\bar{q}}_{LSJ}(k^2)$ the radial wave function of the given $(LSJ)$ channell. Moreover, the momentum-dependent spin factor $R^{SM_S}$ is defined as: $R^{SM_S}(\boldsymbol{k}_\perp, \xi; \nu \bar{\nu}) = \sum_{\nu' \bar{\nu}'} \langle \nu | R^\dagger_M(\boldsymbol{k}_\perp, \xi, m_q) | \nu' \rangle$ $\langle \bar{\nu} | R^\dagger_M(-\boldsymbol{k}_\perp, 1-\xi, m_{\bar{q}}) | \bar{\nu}' \rangle \, \langle \frac{1}{2}\nu' \frac{1}{2}\bar{\nu}' | SM_S \rangle$, where the $2 \times 2$ irreducible representation of the generalized Melosh rotation [3] reads as follows

$$\langle \nu' | R_M(\boldsymbol{k}_\perp, \xi, m) | \nu \rangle = \chi^\dagger_{\nu'} \frac{m + \xi M_0 - i\boldsymbol{\sigma} \cdot (\hat{n} \times \boldsymbol{k})}{\sqrt{(m + \xi M_0)^2 + k_\perp^2}} \chi_\nu \qquad (2)$$

with $\chi_\nu$ being the two-component Pauli spinor and $\boldsymbol{\sigma} \equiv (\sigma_1, \sigma_2, \sigma_3)$ the usual Pauli matrices. Defining $\sigma_0$ as the identity $2 \times 2$ matrix, the spin factor $R^{SM_S}$ can be cast in the following form

$$R^{SM_S}(\boldsymbol{k}_\perp, \xi; \nu \bar{\nu}) = \frac{1}{\sqrt{2}} \sum_{\alpha=0}^{3} a^{SM_S}_\alpha(\boldsymbol{k}_\perp, \xi) \, (\sigma_\alpha)_{\nu \bar{\nu}} \qquad (3)$$

where the quantities $a^{SM_S}_\alpha(\boldsymbol{k}_\perp, \xi)$ are explicitly given by

$$\begin{aligned}
a^{00}_0 &= -k_1(\lambda_q + \lambda_{\bar{q}})/D & , && a^{10}_0 &= ik_2(\lambda_q - \lambda_{\bar{q}})/D \\
a^{00}_1 &= 0 & , && a^{10}_1 &= (\lambda_q \lambda_{\bar{q}} + k_\perp^2)/D \\
a^{00}_2 &= i(\lambda_q \lambda_{\bar{q}} - k_\perp^2)/D & , && a^{10}_2 &= 0 \\
a^{00}_3 &= ik_2(\lambda_q + \lambda_{\bar{q}})/D & , && a^{10}_3 &= -k_1(\lambda_q - \lambda_{\bar{q}})/D \\
a^{1-1}_0 &= (\lambda_q \lambda_{\bar{q}} - k_-^2)/\sqrt{2}D & , && a^{11}_0 &= (\lambda_q \lambda_{\bar{q}} - k_+^2)/\sqrt{2}D \\
a^{1-1}_1 &= -k_-(\lambda_q - \lambda_{\bar{q}})/\sqrt{2}D & , && a^{11}_1 &= k_+(\lambda_q - \lambda_{\bar{q}})/\sqrt{2}D \\
a^{1-1}_2 &= ik_-(\lambda_q + \lambda_{\bar{q}})/\sqrt{2}D & , && a^{11}_2 &= ik_+(\lambda_q + \lambda_{\bar{q}})/\sqrt{2}D \\
a^{1-1}_3 &= -(\lambda_q \lambda_{\bar{q}} + k_-^2)/\sqrt{2}D & , && a^{11}_3 &= (\lambda_q \lambda_{\bar{q}} + k_+^2)/\sqrt{2}D \qquad (4)
\end{aligned}$$

where $D = \sqrt{(\lambda_q^2 + k_\perp^2)(\lambda_{\bar{q}}^2 + k_\perp^2)}$; $\lambda_q = m_q + \xi M_0$; $\lambda_{\bar{q}} = m_{\bar{q}} + (1-\xi)M_0$ and $k_\pm = k_1 \pm ik_2$.

The one-body component of the e.m. current operator is given by

$$\hat{I}_\mu = \sum_{j=q,\bar{q}} \left[ F_1^{(j)}(Q^2)\, \gamma_\mu + F_2^{(j)}(Q^2)\, i\sigma_{\mu\nu}\frac{q^\nu}{2m_j} \right] \quad (5)$$

where $Q^2 = -q \cdot q$ is the squared four-momentum transfer; $F_1^{(j)}(Q^2)$ and $F_2^{(j)}(Q^2)$ are the Dirac and Pauli $CQ$ form factors, respectively (normalized as $F_1^{(q)}(0) = e_q$, $F_2^{(q)}(0) = \kappa_q$, with $e_q$ and $\kappa_q$ being the charge and anomalous magnetic moment of the $CQ$). Within the light-front formalism, all the space-like invariant form factors of a hadron can be determined using only the matrix elements of the *plus* component $I^+$ of the e.m. current operator, evaluated in a frame where $q^+ = 0$; the latter choice allows to suppress the contribution of the pair creation from the vacuum [4]. Thus, using Eqs. (1-5) and performing simple traces involving Pauli matrices, one gets

$$\langle J'M'|\hat{I}^+|JM\rangle = \sum_{j=q,\bar{q}} \sum_{\beta=1,2} F_\beta^{(j)}(Q^2)\, H_\beta^{(j)}(J'M', JM; Q^2) \quad (6)$$

with

$$H_1^{(j)}(J'M', JM; Q^2) = \int dk_\perp d\xi\, \sqrt{A(k_\perp,\xi)A(k'_\perp,\xi)} \quad (7)$$

$$\sum_{SM_S} \tilde{w}_{SM_S}^{JM}(k) \sum_{S'M_{S'}} \tilde{w}_{S'M_{S'}}^{J'M'}(k') \sum_{\alpha=0}^{3} [a_\alpha^{(S'M_{S'})}(k'_\perp,\xi)]^* \, a_\alpha^{(SM_S)}(k_\perp,\xi)$$

$$H_2^{(j)}(J'M', JM; Q^2) = -\frac{Q}{2m_j} \int dk_\perp d\xi\, \sqrt{A(k_\perp,\xi)A(k'_\perp,\xi)} \quad (8)$$

$$\sum_{SM_S} \tilde{w}_{SM_S}^{JM}(k) \sum_{S'M_{S'}} \tilde{w}_{S'M_{S'}}^{J'M'}(k') \{\delta_j [a_0^{(S'M_{S'})}(k'_\perp,\xi)]^* \, a_2^{(SM_S)}(k_\perp,\xi) +$$

$$\delta_j [a_2^{(S'M_{S'})}(k'_\perp,\xi)]^* \, a_0^{(SM_S)}(k_\perp,\xi) - [a_1^{(S'M_{S'})}(k'_\perp,\xi)]^* \, a_3^{(SM_S)}(k_\perp,\xi) +$$

$$[a_3^{(S'M_{S'})}(k'_\perp,\xi)]^* \, a_1^{(SM_S)}(k_\perp,\xi)\}$$

where $\delta_q = i$; $\delta_{\bar{q}} = -i$; $k'_\perp = k_\perp + (1-\xi)q_\perp$ for $j = q$ and $k'_\perp = k_\perp - \xi q_\perp$ for $j = \bar{q}$. Eqs. (6-8) represent our final result, since in terms of them any elastic or transition form factor involving mesons can be calculated. As a matter of fact, we have applied Eqs. (6-8) to the calculation of the elastic form factors of pseudoscalar [5] and vector [6] mesons, as well as of the $\pi\rho$ and $\pi\omega$ [7] transition form factors. In what follows we will focus on the charge form factor of the pion, $F_\pi(Q^2)$, and on the form factor corresponding to the radiative transition $\pi^+ \gamma^* \to \rho^+$, $F_{\pi\rho}(Q^2)$. One has

$$F_\pi(Q^2) = F_1^{(V)}(Q^2) H_1(00,00; Q^2) + F_2^{(V)}(Q^2) H_2(00,00; Q^2) \quad (9)$$

$$F_{\pi\rho}(Q^2) = \frac{2\sqrt{2}}{3Q}[F_1^{(S)}(Q^2) H_1(11,00; Q^2) + F_2^{(S)}(Q^2) H_2(11,00; Q^2)] \quad (10)$$

where $F_{1(2)}^{(S,V)}$ are the isoscalar and isovector parts of the $u$ and $d$ $CQ$ form factors, given by $F_\alpha^{(S)}(Q^2) \equiv 3[F_\alpha^{(u)}(Q^2) + F_\alpha^{(d)}(Q^2)]$ and $F_\alpha^{(V)}(Q^2) \equiv F_\alpha^{(u)}(Q^2) - F_\alpha^{(d)}(Q^2)$. Eqs. (9-10) have been calculated adopting the eigenfunctions of a light-front mass operator, constructed from the effective $q\bar{q}$ Hamiltonian of Ref. [8] and using a simple parametrization for the $CQ$ form factors (see, for more details, [7]). In Fig. 1 the results of our calculations are reported and compared with the predictions of a simple Vector Meson Dominance ($VMD$) model (i.e., $F^{VMD}(Q^2) = 1/(1 + Q^2/M_\rho^2)$, with $M_\rho$ being the $\rho$-meson mass), the results of the Bethe-Salpeter ($BS$) approach of Refs. [9, 10] and those obtained in Refs. [11, 12] using $QCD$ sum rule techniques. It can be seen that the differences among the theoretical calculations of the pion form factor are quite small, so that the existing pion data do not discriminate among various models of the pion structure. On the contrary, the differences among various relativistic calculations of the $\pi\rho$ transition form factor are quite sizeable at $Q^2 > 1\ (GeV/c)^2$; therefore, the measurement of $F_{\pi\rho}(Q^2)$ could help in discriminating among various models of the meson structure.

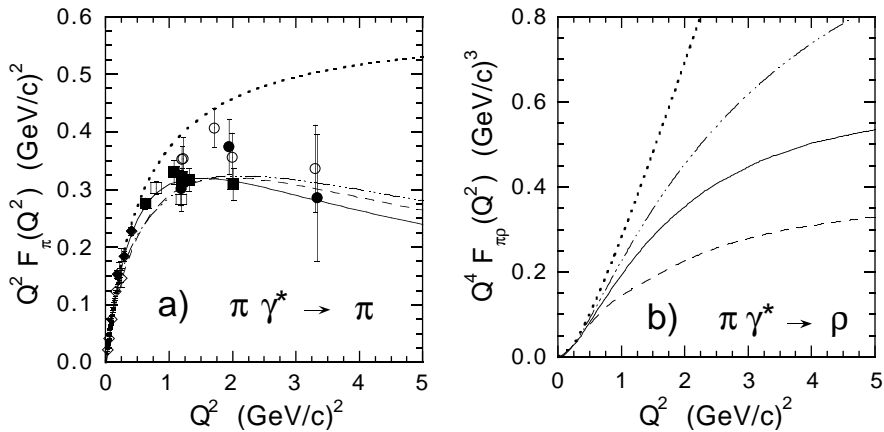

**Figure 1.** a) The elastic form factor of the pion (Eq. (9)), times $Q^2$, vs. $Q^2$. Our results, obtained using for $w_{000}^\pi$ the wave function of Ref. [8] and considering the $CQ$ form factors of Ref. [7], are represented by the solid line. The dotted, dot-dashed and dashed lines correspond to the predictions of a simple $VMD$ model ($\rho$-meson pole only), the $BS$ approach of Ref. [9] and the $QCD$ sum rule technique of Ref. [11], respectively. b) The form factor of the radiative transition $\pi^+ \gamma^* -> \rho^+$ (Eq. (10)), times $Q^4$, vs. $Q^2$. The solid line correspond to our results, obtained using for $w_{000}^\pi$ and $w_{011}^\rho$ the wave functions of Ref. [8] and considering the same $CQ$ form factors as in (a). The dotted, dot-dashed and dashed lines correspond to the predictions of a simple $VMD$ model ($\rho$-meson pole only), the $BS$ approach of Ref. [10] and the $QCD$ sum rule technique of Ref. [12], respectively. (After Ref. [7]).

In conclusion, the elastic and transition electromagnetic form factors of mesons have been investigated within a light-front constituent quark model.

All the relevant formulae needed for the calculation of the space-like matrix elements of the one-body component of the electromagnetic current, including both Dirac and Pauli form factors of the constituent quarks, have been presented. Our results obtained for the pion charge form factor and the $\pi\rho$ transition form factor have been compared with those resulting from various relativistic approaches, showing that the measurements of meson form factors planned at $CEBAF$ could help in discriminating among different models of the meson structure.

**Acknowledgements.**

Two of the authors (I.L.G. and I.M.N.) acknowledge the financial support of the INTAS grant, Ref. No 93-0079.